\documentclass[journal]{IEEEtran}

\usepackage{graphicx} 
\usepackage{amsmath}
\usepackage{xurl}
\usepackage{booktabs} 

\usepackage{caption}  
\usepackage{chngcntr}

\begin{document}

\title{Shape and Substance: Dual-Layer Side-Channel Attacks on Local Vision-Language Models}

\author{
    \begin{minipage}{0.48\textwidth}
        \centering
        Eyal Hadad\\[0.1cm]
        Ben-Gurion University of the Negev, Israel\\
        Faculty of Computer and Information Science\\
        Email: eyalhad@post.bgu.ac.il
    \end{minipage}
    \hfill
    \begin{minipage}{0.48\textwidth}
        \centering
        Mordechai Guri\\[0.1cm]
        Ben-Gurion University of the Negev, Israel\\
        Faculty of Computer and Information Science\\
        Email: gurim@post.bgu.ac.il
    \end{minipage}
}


\maketitle

\begin{abstract}
On-device Vision-Language Models (VLMs) promise data privacy via local execution. However, we show that the architectural shift toward Dynamic High-Resolution preprocessing (e.g., AnyRes) introduces an inherent algorithmic side-channel. Unlike static models, dynamic preprocessing decomposes images into a variable number of patches based on their aspect ratio, creating workload-dependent inputs. We demonstrate a dual-layer attack framework against local VLMs. In Tier 1, an unprivileged attacker can exploit significant execution-time variations using standard unprivileged OS metrics to reliably fingerprint the input's geometry. In Tier 2, by profiling Last-Level Cache (LLC) contention, the attacker can resolve semantic ambiguity within identical geometries, distinguishing between visually dense (e.g., medical X-rays) and sparse (e.g., text documents) content. By evaluating state-of-the-art models such as LLaVA-NeXT and Qwen2-VL, we show that combining these signals enables reliable inference of privacy-sensitive contexts. Finally, we analyze the security engineering trade-offs of mitigating this vulnerability, reveal substantial performance overhead with constant-work padding, and propose practical design recommendations for secure Edge AI deployments.
\end{abstract}

\begin{IEEEkeywords}
Side-channel attacks, Vision-Language Models, Edge AI, Hardware performance counters, Microarchitectural security.
\end{IEEEkeywords}


\section{Introduction}

\IEEEPARstart{T}{he} deployment of Large Language Models (LLMs) is experiencing a significant shift from centralized cloud services to local, on-device execution \cite{zhou2019edge, li2019edge}. A primary motivation for this transition is privacy. Users and organizations are increasingly unwilling to share sensitive personal data, such as medical records, financial documents, and personal photos, with third-party providers. This concern is well-founded, as prior research has demonstrated that cloud-based models can inadvertently memorize and leak training data \cite{carlini2021extracting, shokri2017membership}.
The overall assumption is that running a model locally guarantees data sovereignty. However, if another unprivileged process runs on the same machine, it may still infer sensitive information by observing shared hardware behavior \cite{ristenpart2009hey}.

One might ask why an attacker who is already co-located on the victim’s device would rely on side channels instead of direct screen capture or memory access. The answer lies in modern operating system security models. Contemporary systems enforce strong address-space isolation and memory randomization \cite{shacham2004effectiveness}, which prevents malicious applications from directly reading another process's memory or intercepting its input buffer.

\begin{figure}[t]
    \centering
    \includegraphics[width=\linewidth]{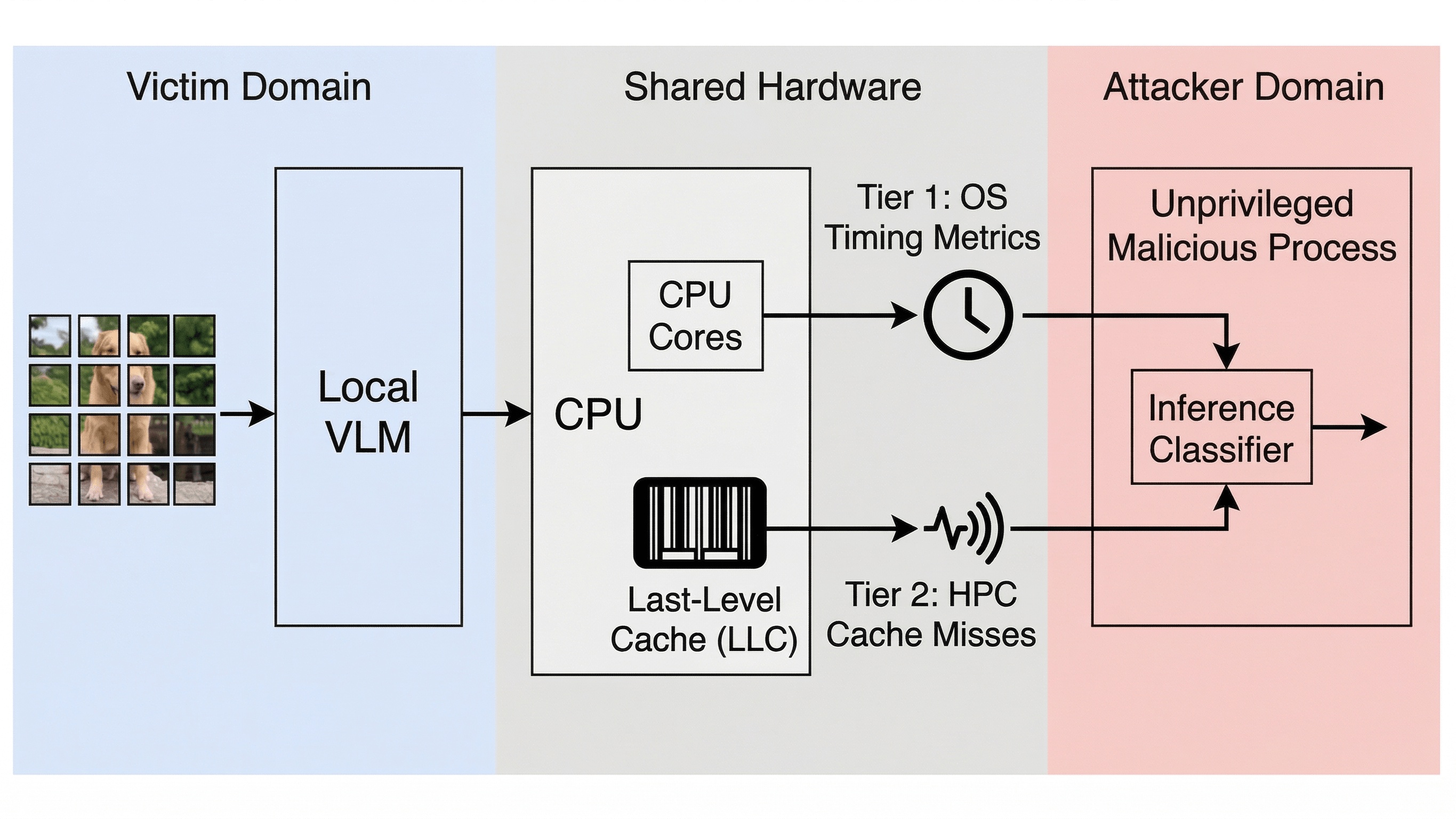} 
    \caption{\textbf{Dual-layer threat model overview.} A co-located, unprivileged process observes shared hardware behavior while a local VLM processes an image with dynamic resolution. The input-dependent workload creates two distinct side-channels: Tier 1 leverages OS-level timing metrics from CPU cores to infer the input's geometric grid (aspect ratio). In contrast, Tier 2 exploits HPC from the LLC to infer the input's visual semantic density.}
    \label{fig:threat_model}
\end{figure}

However, these systems often expose low-level CPU telemetry, specifically hardware performance counters, to user-space processes for debugging and profiling. As a result, while direct access to the input data is blocked, the computational behavior induced by processing that data remains observable through shared hardware resources. This class of leakage has been widely exploited in cryptographic and microarchitectural attacks \cite{ge2018survey, kocher2019spectre, yarom2014flush}.
Figure~\ref{fig:threat_model} illustrates our threat model and the attacker’s two observation channels.

This threat is further amplified by the growing use of unified memory architectures, in which the CPU, GPU, and other accelerators share the same physical memory hierarchy \cite{adiletta2025spill}. In such systems, offloading computation to an accelerator does not eliminate side-channel exposure, as memory access patterns remain observable through shared caches and memory resources \cite{naghibijouybari2018rendered}. As a result, microarchitectural side channels, including those exploiting DRAM addressing \cite{pessl2016drama}, persist even in modern heterogeneous computing platforms.

To understand the threat surface, it is essential to examine the evolution of VLMs. Earlier generations, such as LLaVA v1.5 \cite{liu2024improved}, favored simple and uniform preprocessing by resizing all images to a fixed square resolution (e.g., $336 \times 336$) \cite{dosovitskiy2021image}. While computationally wasteful \cite{schwartz2020green}, this design normalized the computational workload across inputs.

In contrast, recent models such as LLaVA-NeXT \cite{li2024llava} and Qwen2-VL \cite{wang2024qwen2} adopt dynamic high-resolution preprocessing to better preserve visual structure. These mechanisms maintain the original aspect ratio by decomposing an image into a variable number of patches (e.g., $1 \times 2$ for portrait images and $2 \times 2$ for square images). We argue that this shift fundamentally changes image preprocessing from a fixed, input-agnostic operation into an input-dependent control flow, expanding the observable side-channel surface.

We frame this vulnerability not as a specific implementation bug, but as a broader algorithmic side-channel class inherent to input-dependent preprocessing pipelines in Edge AI. As models optimize for efficiency by adapting to the input data, they inevitably create data-dependent execution paths. This fundamental trade-off requires a fresh risk assessment for local AI deployments. 

Specifically, we show that this architectural choice introduces a dual-layer vulnerability. First, at the control-flow level, execution time directly reflects the number of image patches. This allows an attacker to easily infer the underlying grid configuration and the image's aspect ratio. Second, at the memory-access level, the vision encoder’s attention operations \cite{vaswani2017attention} create input-dependent cache activity. Images with high visual complexity cause significantly more cache contention than sparse, text-like documents \cite{liu2015last}. By combining timing information with cache behavior, an attacker can resolve the ambiguity of a single metric and extract both geometric and semantic information.

Our evaluation demonstrates that this dynamic preprocessing design creates a practical security risk, enabling detailed context inference:
\begin{itemize}
    \item \textbf{Aspect Ratio Inference:} Execution-time variations in dynamic models, such as Qwen2-VL, allow reliable inference of image dimensions due to substantial latency differences across grid configurations.
    \item \textbf{Semantic Context Inference:} By jointly analyzing timing and cache activity, the attack distinguishes between different high-level input contexts, including visually dense and sparse scenarios, with high recall.
    \item \textbf{Structural vs. Hardware-Dependent Leakage:} We show that geometric leakage arises from the dynamic preprocessing design itself, while semantic leakage depends on hardware characteristics and is mitigated on systems with larger LLC \cite{ge2018survey}.
\end{itemize}

\section{Background}

\subsection{Local Vision-Language Models}
\label{subsec:local_vlms}
A VLM is composed of multiple components rather than a single monolithic model. Typically, it consists of three main stages:
\begin{enumerate}
    \item A \textbf{Vision encoder} (e.g., CLIP \cite{radford2021learning} or SigLIP \cite{tschannen2025siglip}) that converts the input image into visual features, typically utilizing a Vision Transformer backbone \cite{dosovitskiy2021image}.
    \item A \textbf{Projector} (or adapter) that maps visual features into a representation compatible with the language model, as used in architectures such as BLIP-2 \cite{li2023blip} and Flamingo \cite{alayrac2022flamingo}.
    \item An \textbf{LLM} that jointly processes the projected visual features and text instructions to generate a response \cite{liu2024improved}.
\end{enumerate}

In cloud-based deployments, such as GPT-4V \cite{achiam2023gpt}, these components may execute on separate or dedicated hardware. In contrast, during local on-device inference on personal systems such as laptops or workstations, the entire pipeline runs on a single physical device. As a result, the vision encoder and the language model share the same memory hierarchy, including LLC. In our experiments, we consider common local inference setups, such as implementations based on \texttt{llama.cpp} \cite{llamacpp}, which optimize the LLaMA architecture for consumer hardware.

\subsection{Dynamic Resolution (AnyRes)}
\label{subsec:anyres}

To understand the root cause of the side-channel, it is necessary to examine how modern VLMs process image inputs. Earlier models, such as LLaVA v1.5 \cite{liu2024improved}, relied on static preprocessing, where all images were resized or padded to a fixed square resolution (e.g., $336 \times 336$), following standard computer vision practices \cite{he2016deep}. This approach resulted in largely input-agnostic computational workloads.

\begin{figure}[t] 
    \centering
    \includegraphics[width=\linewidth]{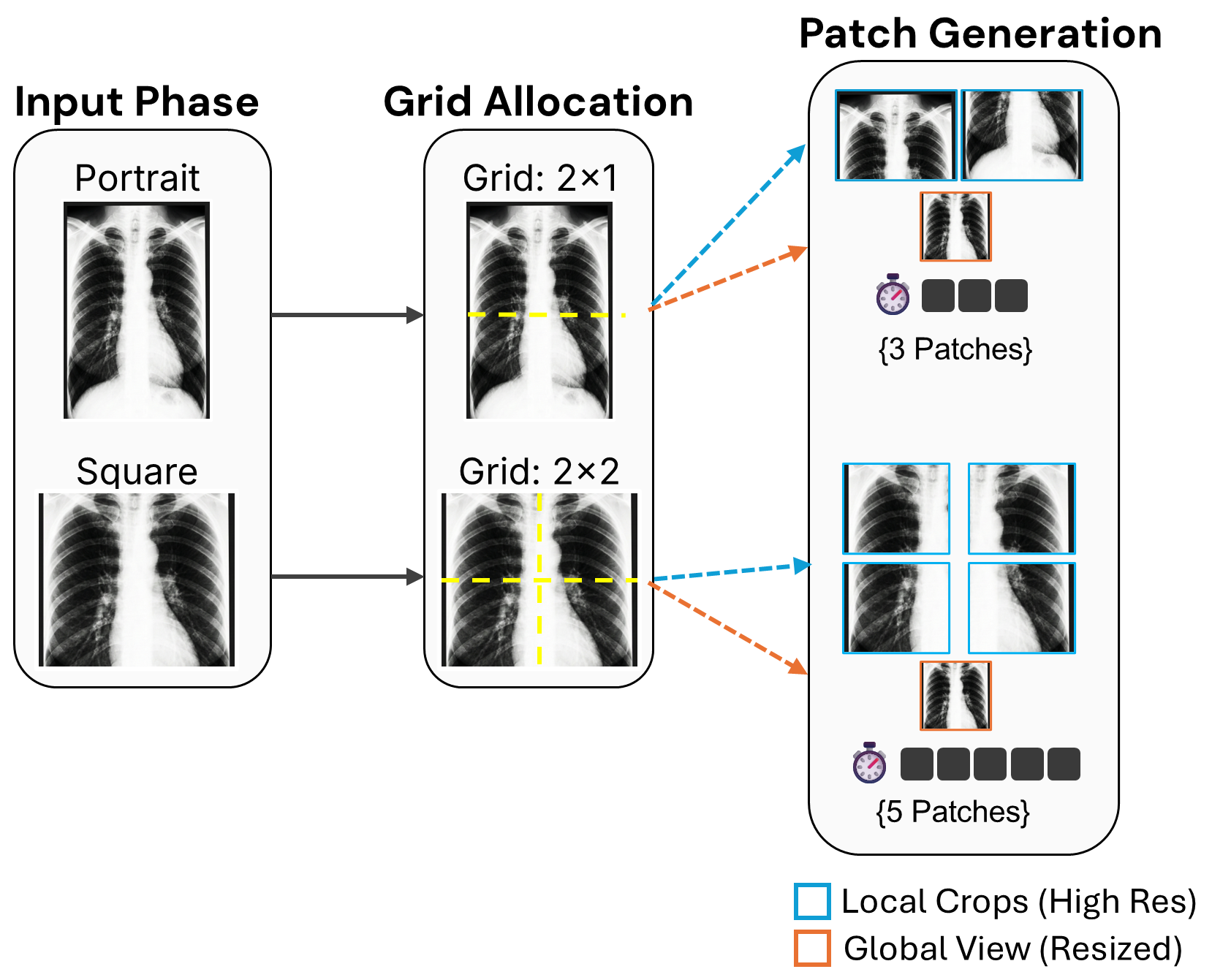}
   \caption{\textbf{AnyRes dynamic preprocessing.} The model tiles the image into an $(m\times n)$ grid based on aspect ratio and adds a global view, yielding $(m\times n)+1$ patches. Different grids (e.g., $1\times2$ vs. $2\times2$) produce different amounts of work, creating a timing signal.}
    \label{fig:mechanism}
\end{figure}

In contrast, recent architectures such as LLaVA-NeXT \cite{li2024llava} and Qwen2-VL \cite{wang2024qwen2} adopt dynamic high-resolution preprocessing strategies, commonly referred to as AnyRes. Inspired by architectures like NaViT \cite{dehghani2024patch}, rather than enforcing a fixed input shape, these models adapt to the image’s original aspect ratio using a grid-based decomposition. Figure~\ref{fig:mechanism} illustrates how AnyRes maps aspect ratios to different grid configurations and patch counts.

To preserve global context that may be lost when processing high-resolution local crops (e.g., observing only part of an object), AnyRes pipelines typically include an additional downsampled global view alongside the local patches. For a given input image, the model selects a grid configuration $(m \times n)$ that minimizes aspect ratio distortion, resulting in a total number of processed patches given by
\[
N_{\text{total}} = (m \times n) + 1.
\]

For example, a vertical smartphone image commonly triggers a $1 \times 2$ grid, yielding three patches (two local crops and one global view), whereas a square document triggers a $2 \times 2$ grid, yielding five patches. Since the vision encoder processes each patch either sequentially or in batches, utilizing optimization techniques like FlashAttention \cite{dao2022flash}, square inputs incur a higher computational workload than portrait inputs, effectively creating an exploitable algorithmic complexity gap \cite{crosby2003denial}.

\subsection{Hardware Performance Counters}
\label{subsec:hpc}
To bridge the gap between this algorithmic complexity and observable physical phenomena, we rely on microarchitectural telemetry. Hardware Performance Counters (HPCs) allow software to monitor low-level architectural events during program execution. Since foundational research on timing attacks \cite{kocher1996timing} and cache-based side channels \cite{yarom2014flush}, it has been well established that such measurements can leak information about a victim’s execution behavior.

On Linux systems, hardware counters are accessible through the \texttt{perf\_event\_open} interface. While modern operating systems enforce strong isolation mechanisms that prevent direct access to another process’s memory, they often permit user-space access to performance counters for profiling and debugging. Accordingly, our threat model assumes a co-located attacker running an unprivileged process that monitors hardware events under default system configurations, such as those where \texttt{perf\_event\_paranoid} $\leq 2$.

As described in Section~\ref{subsec:anyres}, dynamic image preprocessing introduces input-dependent variation in computational workload. We show that this variation induces a distinct and measurable signal in execution time and cache activity, which can be exploited by an unprivileged attacker to infer properties of the visual input \cite{yan2020cache}.

In our experiments, we use the Linux \texttt{perf} subsystem to collect hardware performance counters. While certain controls, such as clearing the cache via \texttt{/proc/sys/vm/drop\_caches}, require root privileges for reproducibility, accessing hardware counters themselves is available to unprivileged user-space processes in default Linux configurations.

\section{Related Work}
\label{sec:related_work}
Our attack relies on the hardware behaviors described in the previous section. In this section, we compare our approach with existing model extraction and side-channel techniques.

\subsection{Model Extraction vs. Physical Leakage}
A significant body of research focuses on stealing machine learning models or extracting sensitive training data via Application Programming Interfaces (APIs). Techniques such as Model Inversion \cite{fredrikson2015model} or Membership Inference \cite{shokri2017membership} rely on extensive model querying to reconstruct data from its outputs and confidence scores. In contrast, our work explores physical, passive microarchitectural leakage. While prior studies have successfully demonstrated the extraction of DNN architectures and hyperparameters via hardware side-channels \cite{zhang2021stealing}, we expand this threat paradigm to the inference data itself. We demonstrate that an attacker does not need API access to the model's outputs. Instead, they can infer properties of the private inputs solely by observing shared hardware resources during local execution.

\subsection{Text-Only LLMs vs. Multimodal Encoders}
Recent work has shown that LLMs are vulnerable to side-channel attacks. For instance, previous studies have demonstrated that hardware cache behavior can leak information during the token generation phase of local LLMs \cite{gao2025know}. Other studies have exploited timing variations in cloud-based serving systems or prompt caching \cite{zheng2024inputsnatch, song2025early, gu2025auditing}, or targeted specific architectural choices, such as Mixture-of-Experts \cite{ding2025moecho}. However, these attacks focus solely on the text-decoding phase. Our research uniquely targets the vision encoder in multimodal pipelines, demonstrating that processing image data introduces new vulnerabilities before any text is generated.

\subsection{Privacy Leaks in User Workloads}
Inferring user context from hardware performance is a known threat in applied systems security. Prior research has shown that graphics rendering pipelines and browser activities can leak sensitive information \cite{naghibijouybari2018rendered}. For example, attackers can use cache-based side channels to recover keystrokes from graphics libraries \cite{wang2019keystrokes, kotcher2013cross} or identify encrypted video streams and websites based on traffic and memory patterns \cite{cai2012touching, schuster2017beauty}. Moreover, recent advances highlight the potency of leveraging LLC telemetry and hardware events to model and detect complex microarchitectural patterns \cite{kim2023deep}, reinforcing the viability of cache activity as a rich semantic signal. Our findings extend this threat model to local AI applications, showing that structural visual complexity translates into a highly reliable side-channel signal.

\subsection{Our Positioning}
Our work introduces a new perspective on AI privacy by bridging multimodal architectures with systems security. Specifically, we contribute the following:
\begin{itemize}
    \item \textbf{A New Leakage Class:} We identify dynamic high-resolution preprocessing (e.g., AnyRes) as an inherent algorithmic side-channel, shifting the focus from simple text generation to input-dependent vision pipelines.
    \item \textbf{Dual-Layer Attack Vectors:} We combine a zero-privilege timing attack (geometry inference) with a cache-based attack (semantic inference) to overcome the limitations of single-metric observations.
    \item \textbf{Risk Assessment for Edge AI:} We provide practical evidence that running VLMs locally does not guarantee data privacy, highlighting a systematic vulnerability arising from input-dependent preprocessing in modern edge AI deployments.
\end{itemize}

\section{Threat Model}
\label{sec:threat_model}
Building on these concepts, we now define the formal threat model for our attack, detailing the attacker's capabilities and the assumptions about the victim's environment.

\subsection{Attacker Capabilities}
\label{subsec:attacker}

We define the attacker as a malicious native application running in user space on the victim's device. To present a realistic scenario, we assume the attacker runs under the same user account (UID) as the victim but operates in a separate process. The operating system enforces standard address-space isolation, meaning the attacker cannot directly access the victim's memory, read the victim's input images, or capture the victim's screen. Furthermore, the attacker is unprivileged and lacks root (administrator) access.

Because the attacker shares the same physical hardware as the VLM process, we structure our threat model into two distinct tiers of observation capabilities:

\begin{enumerate}
    \item \textbf{Tier 1: Coarse Timing Observation (Zero-Privilege):} The leakage in our system causes large differences in inference execution time, often on the order of tens of seconds. As a result, an attacker does not need access to hardware performance counters. Instead, they can simply measure the inference phase duration using standard OS-level metrics, such as CPU utilization or wall-clock time (e.g., via \texttt{/proc/stat} on Linux). This attack requires no special privileges and remains possible even if access to performance telemetry is restricted.

    \item \textbf{Tier 2: Microarchitectural Profiling (perf):} To infer semantic information about the processed input, the attacker monitors hardware performance counters, particularly LLC misses, using the \texttt{perf\_event\_open} interface. Although some Linux systems restrict access to these counters (e.g., when \texttt{perf\_event\_paranoid} is set to 3 or higher), our target users—AI developers and researchers running local VLMs—often enable performance profiling for debugging and optimization. In such environments, \texttt{perf\_event\_paranoid} is commonly set to $\leq 2$, allowing unprivileged processes to access these counters and exposing the system to our semantic side-channel.
\end{enumerate}

The attacker's combined goal is to use Tier 1 (timing) to determine the image's geometry and Tier 2 (cache contention) to resolve its sensitive semantic context.

\subsection{Victim Assumptions}
\label{subsec:victim}
The victim is a user or organization utilizing a VLM locally to process sensitive data. The primary motivation for local deployment is privacy, ensuring that data does not leave the physical device.

We assume the following about the victim’s environment:
\begin{itemize}
    \item \textbf{Software Stack:} The victim uses widely adopted inference frameworks, such as open-source implementations based on \texttt{llama.cpp} or PyTorch, to run the VLM.
    \item \textbf{Hardware:} The victim operates on standard consumer or workstation hardware (e.g., Intel or AMD CPUs). While the system may include a GPU, the CPU remains actively involved in input preprocessing, memory management, and control flow decisions of dynamic resolution mechanisms. As a result, shared microarchitectural resources remain a source of information leakage \cite{yarom2014flush}.
    \item \textbf{Data Sensitivity:} The input images contain private information, such as medical imagery (e.g., X-rays), financial documents, or personal identification, for which geometric properties and visual complexity may correlate with the underlying content.
\end{itemize}

\section{Deterministic Leakage}
We begin our evaluation by validating the first tier of our threat model: the zero-privilege timing attack. In this section, we aim to prove that the algorithmic control-flow introduced by dynamic preprocessing produces a deterministic and easily observable timing signal.
\label{sec:deterministic}

\begin{table}[t!]
    \centering
    \small
    \renewcommand{\arraystretch}{1.2}
    \resizebox{\columnwidth}{!}{
    \begin{tabular}{l l l}
        \toprule
        \textbf{Component} & \textbf{Intel Setup} & \textbf{AMD Setup} \\
        \midrule
        \textbf{Experimental Role} & Primary Evaluation & Cross-Platform Validation \\
        \textbf{CPU} & Core i7-13700 (16 Cores) & Ryzen 9 7950X (16 Cores) \\
        \textbf{LLC} & 30 MB & 64 MB \\
        \textbf{OS Kernel} & Linux 6.8 (Ubuntu 24.04) & Linux 6.8 (Ubuntu 24.04) \\
        \bottomrule
    \end{tabular}
    }
    \vspace{1pt}
    \caption{\textbf{Hardware Specifications.} The Intel setup serves as the primary reference platform, while the AMD setup is used to validate architectural robustness under a significantly larger LLC.}
    \label{tab:specs}
\end{table}

\begin{figure*}[t!]
    \centering
    \includegraphics[width=0.8\textwidth]{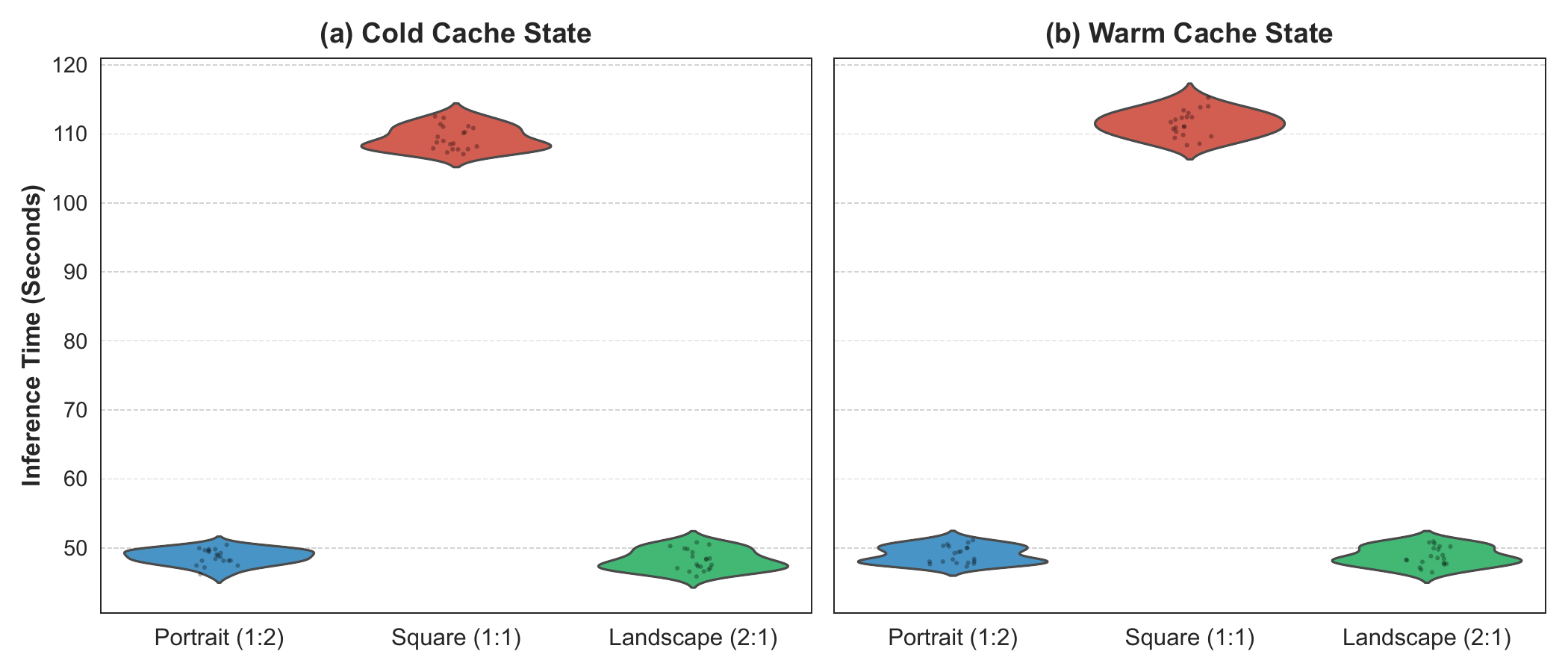} 
    \caption{\textbf{Deterministic Geometric Leakage.} Distribution of inference time across aspect ratios in (a) Cold Cache and (b) Warm Cache states. Note the distinct "pyramid" structure: Square inputs (1:1) incur a $\sim 2.3\times$ latency penalty compared to rectangular inputs (1:2, 2:1), creating non-overlapping clusters that enable 100\% classification accuracy regardless of cache state.}
    \label{fig:aspect_ratio_violin}
\end{figure*}

\subsection{Experimental Setup}
\label{subsec:setup}

We evaluated the vulnerability on two distinct hardware configurations representing high-end consumer workstations (Table~\ref{tab:specs}). The Intel Core i7-13700 served as the primary reference platform for the deterministic and semantic leakage analysis. The AMD Ryzen 9 7950X was used to validate robustness across microarchitectural designs.

Unless otherwise noted, all reported measurements were collected on the Intel setup running Ubuntu 24.04 LTS with the \texttt{llama.cpp} inference engine~\cite{llamacpp} (for full reproducibility details, including core pinning and software versions, see Appendix A in the supplementary material).

\subsection{Experimental Methodology}
To isolate side-channel leakage induced by dynamic resolution preprocessing, we constructed a synthetic dataset containing $N=250$ unique images per category. The dataset consists of two primary subsets:
\begin{itemize}
    \item \textbf{Geometric Benchmarks:} Images with controlled aspect ratios (square 1:1 vs. portrait 1:2), designed to trigger distinct grid configurations and isolate geometry-driven execution time.
    \item \textbf{Semantic Benchmarks:} Content-specific categories, including documents (sparse text), X-rays (dense structure), crypto-noise (randomized patterns), and synthetic scenes designed to mimic the visual characteristics of natural landscapes (e.g., color gradients and structural edges). For brevity, we refer to this synthetic category as Nature in the following analysis. (generation scripts available in the repository). 
\end{itemize}

This dataset design allows us to isolate both algorithmic control-flow effects (via the geometric benchmarks) and memory-access behavior (via the semantic categories), while maintaining controlled visual characteristics (see Appendix B in the supplementary material).

To establish precise ground truth for the algorithmic vulnerability, we monitored the prompt evaluation time reported by the inference engine, isolating the visual processing workload. While an unprivileged attacker cannot access these internal logs, this preprocessing phase consumes the majority of the overall execution time. As we demonstrate later (see Section \ref{sec:deterministic}), the temporal differences in this phase are vast enough that standard end-to-end wall-clock timing captures the same exploitable signal without requiring internal metrics.

To assess robustness under different system conditions, experiments were conducted in two cache states:
\begin{enumerate}
    \item \textbf{Cold Cache:} System caches were flushed prior to each run using \texttt{drop\_caches} to enforce a clean execution state.\footnote{\textbf{Privileges:} Root access was used solely to enforce a deterministic cold-cache condition for experimental reproducibility. The attack itself runs entirely in user space and relies only on observable execution time.}
    \item \textbf{Warm Cache:} The model was executed repeatedly without clearing caches, simulating a typical interactive usage scenario.
\end{enumerate}

\subsection{Analysis of Execution Time}
To isolate the contribution of the vision encoder from the rest of the inference pipeline, we use the \textbf{prompt evaluation time} reported by the inference engine logs as ground truth. \footnote{
\textbf{Measurement methodology.}
For experimental reproducibility, we collect measurements using the standard Linux \texttt{perf} subsystem. 
However, the attack described in this section does not require access to performance counters and relies only on observable execution time. Although our threat model assumes an attacker using the \texttt{perf\_event\_open} interface, the \texttt{perf} utility provides a stable and widely used mechanism for collecting performance counter data.}
This metric captures the time spent encoding image patches and projecting them into the language model’s embedding space, excluding system initialization overhead.

\textbf{Attacker Perspective:} While a real-world attacker cannot access internal inference logs, they can observe the total wall-clock duration of the high-CPU-load interval during image processing. Our measurements confirm that the prompt evaluation phase dominates overall execution time, accounting for more than $95\%$ of the duration. Consequently, the timing differences reported here are directly observable by an unprivileged attacker using standard timing measurements, a standard assumption in microarchitectural timing attacks \cite{ge2018survey}.

As shown in Figure~\ref{fig:aspect_ratio_violin}, inference time exhibits a clear and consistent separation across aspect ratios. Square inputs (1:1) incur substantially higher latency than rectangular inputs (portrait or landscape), resulting in well-separated timing distributions.

Specifically, square inputs require approximately 110 seconds on average, corresponding to the processing of five patches (four local patches and one global view). In contrast, rectangular inputs complete in approximately 48 seconds, reflecting the processing of only three patches. This yields a latency ratio of approximately $\times$2.3.

Importantly, this separation persists across cache states. Although absolute execution times are slightly lower in the warm-cache setting, the relative gap and non-overlapping clusters remain unchanged, indicating that caching effects do not mitigate this architectural side-channel.

\section{Semantic Inference}
\label{sec:semantic}

\subsection{Motivation for Semantic Leakage}
\label{subsec:semantic_motivation}

Having shown the feasibility of the Tier 1 zero-privilege geometry attack, we now explore the Tier 2 microarchitectural attack. The aspect-ratio leakage described in Section~\ref{sec:deterministic} arises from deterministic control-flow decisions in dynamic preprocessing. In this section, we examine whether additional information can be inferred from the visual content itself, even when input images share identical dimensions and therefore trigger the same preprocessing grid.

Modern vision encoders rely on attention-based architectures whose memory access patterns depend on the structure and complexity of the input \cite{dao2022flash}. Images with dense visual structure, such as medical imagery, induce different memory access behavior than sparse inputs, such as scanned documents with large uniform regions. While these differences may not significantly affect total execution time, they can manifest as variations in cache usage and memory contention.

We therefore hypothesize that visual content introduces a secondary, probabilistic side-channel that complements the deterministic geometric leakage. In the following experiments, we empirically evaluate whether this signal is observable through hardware performance counters and whether it can be exploited to infer semantic properties of the input image.

\subsection{The Single-Dimension Limit}
\label{subsec:raw_metrics}
To evaluate whether semantic information can be inferred independently of image geometry, we analyzed raw hardware metrics for $N=250$ images of identical dimensions ($672 \times 672$) but varying visual content. Table~\ref{tab:semantic_means} reports the mean execution time and LLC miss counts for each content class under a warm-cache configuration.

\textbf{Execution Time:} As expected, execution time is a strong discriminator for aspect ratio but loses predictive power when the geometry is fixed. Since all inputs trigger the same $2 \times 2$ preprocessing grid, the relative execution-time difference between sparse documents and dense X-ray images is below $3\%$, rendering it indistinguishable from normal system variability.

\textbf{LLC Miss Behavior:} In contrast, LLC miss counts reveal a consistent pattern across content classes. Noise-like inputs exhibit the lowest cache contention, while visually dense and structured images induce higher cache pressure.

\begin{itemize}
    \item \textbf{Lowest Contention (Noise):} Randomized, noise-like images produce the lowest LLC miss counts ($1.69 \times 10^{10}$), suggesting relatively uniform memory access. Although randomness typically increases cache misses, attention-based vision encoders fail to extract meaningful features from noise, resulting in smaller active working sets. This aligns with findings linking input entropy to observable side-channel leakage boundaries \cite{zhang2023guessing}.
    \item \textbf{Highest Contention (Dense Images):} Structured visual content, such as X-rays and natural scenes, induces higher LLC miss rates ($\approx 1.79 \times 10^{10}$). Dense structures force the attention mechanism to actively extract and process a larger number of features across the spatial grid, increasing memory pressure and LLC contention.
\end{itemize}

Representative visual examples across classes are provided in Appendix B of the supplementary material.

\begin{table}[t]
    \centering
    \small
    \renewcommand{\arraystretch}{1.2}
    
    \begin{tabular}{l | c c}
        \toprule
        \textbf{Content Class} & \textbf{Time (s)} & \textbf{LLC Misses ($10^9$)} \\
        \midrule
        Crypto-Noise & 107.4 & $\mathbf{16.9}$ \textit{(Lowest)} \\
        Document     & 107.6 & $17.6$ \\
        Nature       & 109.0 & $17.8$ \\
        X-Ray        & 110.5 & $\mathbf{17.9}$ \textit{(Highest)} \\
        \bottomrule
    \end{tabular}
    \vspace{2pt}
    \caption{\textbf{Mean hardware metrics (warm cache).} Execution time remains nearly constant across content classes, while LLC misses exhibit a consistent semantic ordering.}
    \label{tab:semantic_means}
\end{table}

\subsection{The Ambiguity Challenge}
While Table~\ref{tab:semantic_means} reveals a measurable divergence between classes, relying on a single hardware metric is insufficient for robust single-shot inference. Microarchitectural state is inherently noisy due to dynamic OS memory management, background interrupts, and non-deterministic cache replacement policies. Consequently, the LLC miss distributions of visually dense classes exhibit non-negligible overlap. This inherent system variance prevents the isolation of semantic context using cache telemetry alone, motivating our dual-layer approach.

These limitations highlight the need for a multi-dimensional approach that combines complementary signals. In the following section, we demonstrate that jointly analyzing execution time and cache behavior resolves this ambiguity and enables reliable semantic inference.

\section{The Combined Attack}
\label{sec:combined_attack}

\subsection{The Multi-Dimensional Attack Vector}
In Section~\ref{sec:semantic}, we established that while LLC Misses can indicate semantic density, using this metric alone is inherently noisy. Similarly, Section~\ref{sec:deterministic} showed that execution time creates broad, distinct clusters based on aspect ratio, but it cannot resolve content differences within those clusters because inputs of the same geometry incur almost identical processing times.

To overcome these limitations, we propose a \textbf{Combined Attack} that models the leakage as a two-dimensional vector $\vec{v} = (t, c)$, where $t$ denotes execution time and $c$ denotes LLC miss activity.

\subsection{Attack Methodology}
To contextualize this combined attack within realistic environments, we translate our base synthetic categories into specific high-risk assets. Unlike the isolated semantic evaluation in Section~\ref{sec:semantic}, where all images were fixed to a square geometry, this dataset assigns both a characteristic aspect ratio and a structural density to each class.

Specifically, we map the base image generation functions to real-world proxies as follows:
\begin{itemize}
    \item \textbf{Medical Report:} Generated using the sparse text function, formatted as a Portrait ($1:2$).
    \item \textbf{Chest X-Ray:} Generated using the dense structural function, formatted as a Portrait ($1:2$).
    \item \textbf{Encrypted Data:} Generated using the random noise function, formatted as a Square ($1:1$). As established earlier, noise fails to activate the attention mechanism, resulting in low cache contention.
    \item \textbf{Technical Schematic:} Generated using the dense structural function, formatted as a Square ($1:1$) to proxy highly detailed engineering diagrams.
\end{itemize}

By creating these physical combinations, we establish two representative real-world scenarios to evaluate the effectiveness of the attack:
\begin{enumerate}
    \item \textbf{Scenario A:} Distinguishing \textit{Medical Reports} (low cache contention) from \textit{Chest X-Rays} (high cache contention) within the Portrait geometry cluster.
    \item \textbf{Scenario B:} Distinguishing \textit{Encrypted Data} (low cache contention) from \textit{Technical Schematics} (high cache contention) within the Square geometry cluster.
\end{enumerate}

\subsubsection{Visualizing the Joint Feature Space}
\begin{figure}[h]
    \centering
    \includegraphics[width=\linewidth]{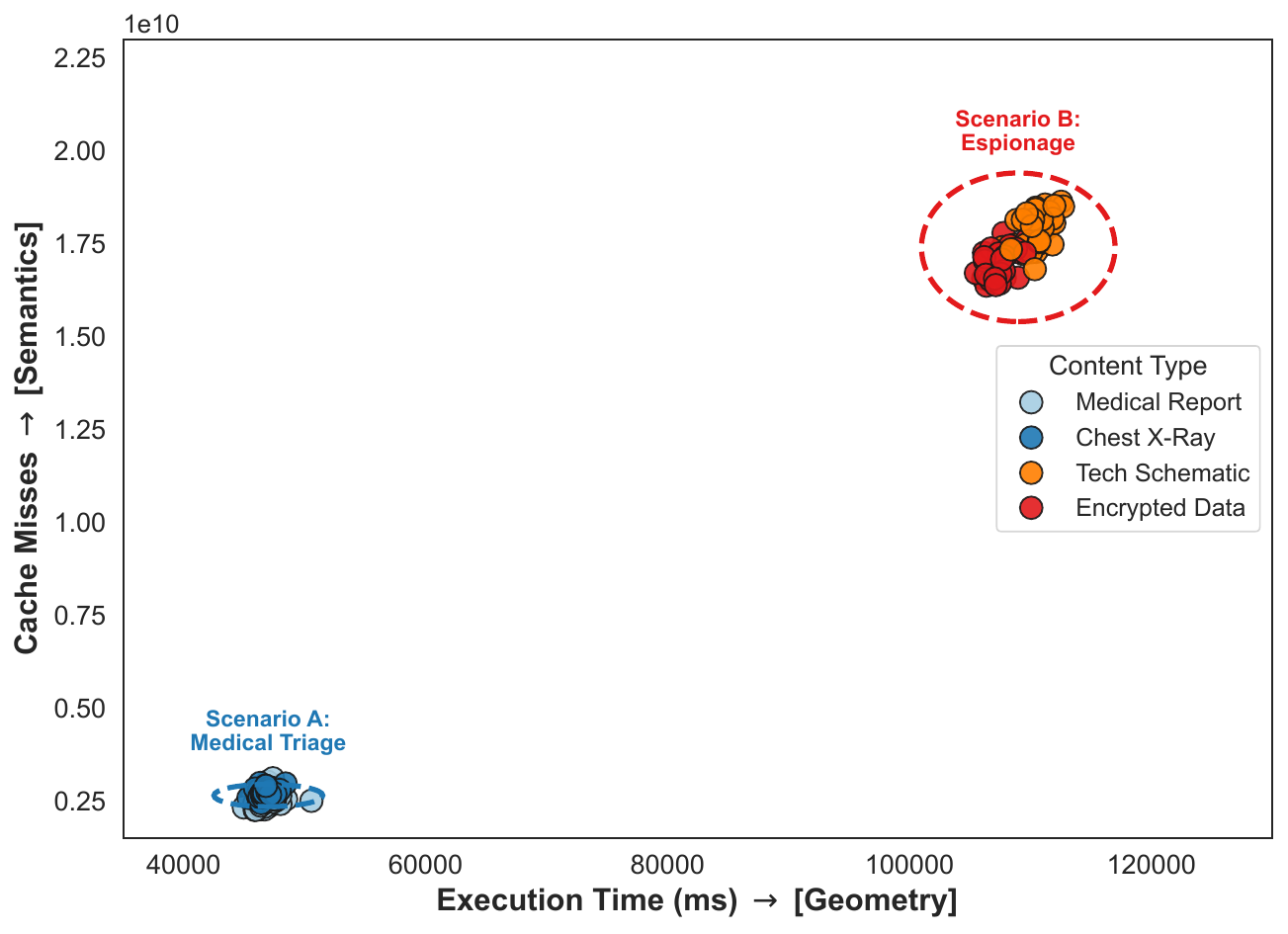}
    \caption{\textbf{Joint timing and cache leakage.} Projection of the Combined Attack vector into a two-dimensional space. The X-axis (execution time) separates inputs by geometry clusters (portrait vs. square) due to varying patch counts. The Y-axis (LLC misses) resolves the semantic ambiguity within each geometric cluster (e.g., separating dense X-ray images from sparse documents).}
    \label{fig:combined_scatter}
\end{figure}

Figure~\ref{fig:combined_scatter} visualizes these scenarios by projecting the collected data into the proposed two-dimensional vector space. The physical behavior of the dynamic VLM perfectly explains the resulting distribution: 
First, the X-axis (Execution Time) cleanly separates the inputs into two distant clusters. This massive gap occurs because the Portrait inputs (Scenario A) are decomposed into three patches, whereas the Square inputs (Scenario B) are decomposed into five patches. However, within each isolated cluster, the data points overlap horizontally, confirming that timing alone cannot resolve semantic content. 
Second, the Y-axis (LLC Misses) resolves this ambiguity. Within each geometry cluster, the data points separate vertically based on visual complexity, allowing an attacker to distinguish dense structures from sparse ones.

\subsubsection{Dataset and Protocol}
We constructed a dedicated dataset spanning these four content combinations. The dataset was split into training ($70\%$) and testing ($30\%$) sets using stratified sampling to preserve class balance.

\subsubsection{Model Choice: Interpretable Baseline Classifier}
Our rationale is that if a simple and interpretable classifier can achieve strong performance using only timing and cache features, this indicates that the leakage is systematic rather than arising from incidental noise. The use of a shallow decision tree (maximum depth of 3) further enables inspection of the learned decision boundaries, highlighting the separability of the leakage signals without relying on complex or black-box models, aligning with principles of explainable AI \cite{ribeiro2016should}.

\begin{table}[t]
    \centering
    \small
    \renewcommand{\arraystretch}{1.2}
    \resizebox{\columnwidth}{!}{
    \begin{tabular}{l | r r | r r}
        \toprule
        & \multicolumn{2}{c|}{\textbf{Intel (30MB)}} & \multicolumn{2}{c}{\textbf{AMD (64MB)}} \\
        \textbf{Evaluation Phase} & \textbf{Low} & \textbf{High} & \textbf{Low} & \textbf{High} \\
        \midrule
        \textbf{Time (s)} & 49 & 111 & 18 & 29 \\
        \textit{(Portrait vs. Square)} & \multicolumn{2}{c|}{\textit{2.27x Gap}} & \multicolumn{2}{c}{\textit{1.61x Gap}} \\
        \midrule
        \textbf{LLC Misses ($10^9$)} & 16.9 & 17.9 & 4.6 & 4.6 \\
        \textit{(Noise vs. X-Ray)} & \multicolumn{2}{c|}{\textit{Observable}} & \multicolumn{2}{c}{\textit{Not Observable}} \\
        \bottomrule
    \end{tabular}
    }
    \vspace{1pt}
    \caption{\textbf{Cross-Architecture Evaluation.} The geometric timing gap remains clear across platforms, reflecting algorithmic control-flow. Conversely, the semantic cache signal is not observable on the AMD architecture due to its larger LLC capacity.}
    \label{tab:cross_arch_results}
\end{table}

\subsection{Experimental Results}
\begin{figure}[t]
    \centering
    \includegraphics[width=\linewidth]{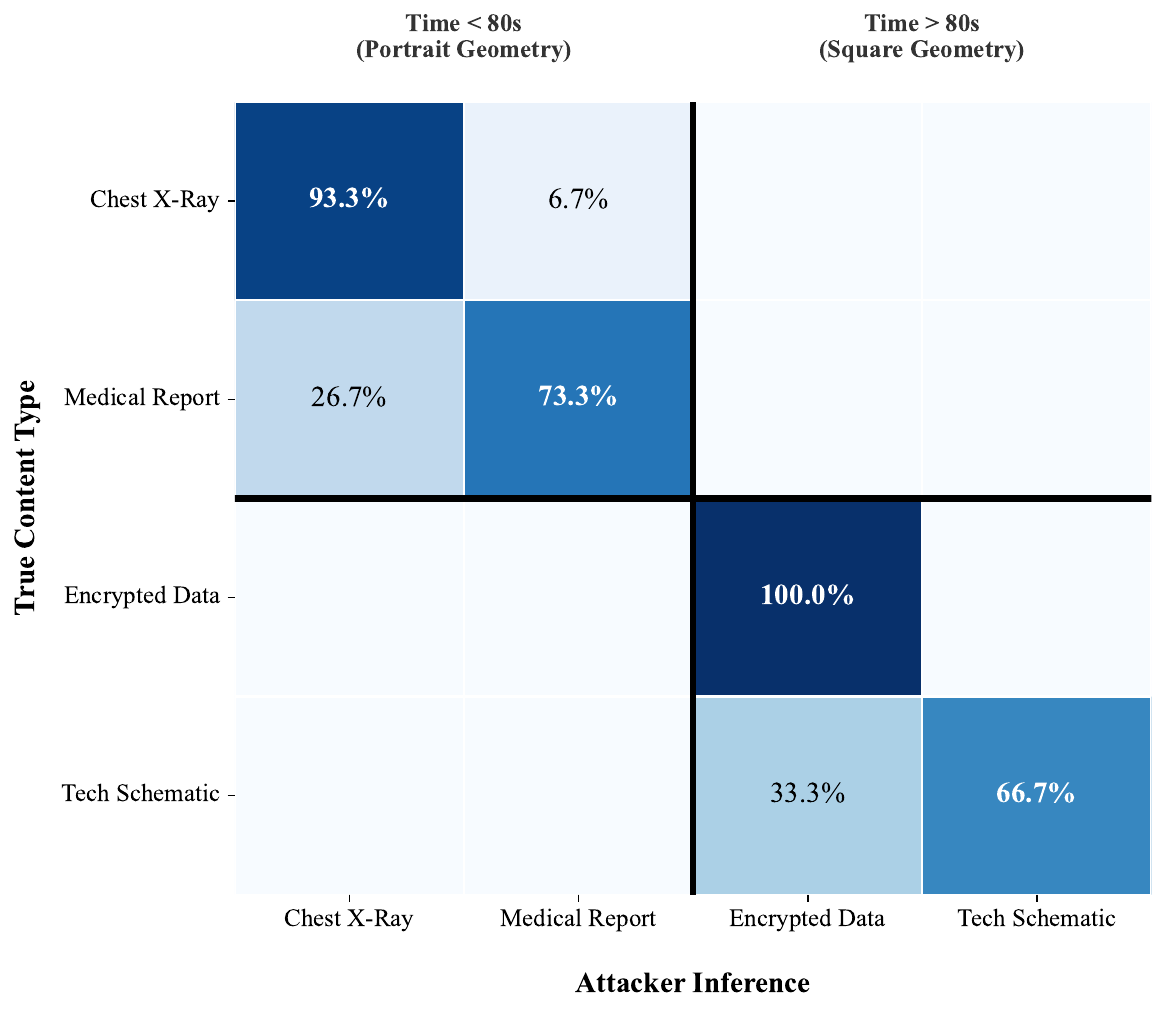}
    \caption{\textbf{Confusion Matrix of the Combined Attack.} The results visually validate the dual-layer threat model: deterministic timing differences eliminate cross-geometry errors (empty off-diagonal quadrants), while cache-miss telemetry resolves semantic ambiguity within identical geometries. Notably, the attack achieves perfect or near-perfect classification for privacy-critical targets, such as encrypted data and X-ray images.}
    \label{fig:confusion_matrix}
\end{figure}

The model achieved an overall accuracy of 0.84 on the test set. However, a closer look at the performance metrics (Table~\ref{tab:classification_report}) and the prediction distribution (Figure~\ref{fig:confusion_matrix}) helps explain how the attack works and why it raises privacy concerns.

First, as shown in Figure~\ref{fig:confusion_matrix}, the model never confuses images with different aspect ratios. This shows that the first signal (the overall inference time) separates images by their geometric layout in a fully deterministic manner. In other words, the timing information alone reliably reveals the input's grid structure.

The second signal comes from LLC activity. Unlike the deterministic timing, this microarchitectural signal introduces small variations. However, it is essential for resolving semantic ambiguity, allowing the model to distinguish between visually dense and sparse content within the same geometry. 

\begin{figure}[t]
    \centering
    \includegraphics[width=\linewidth]{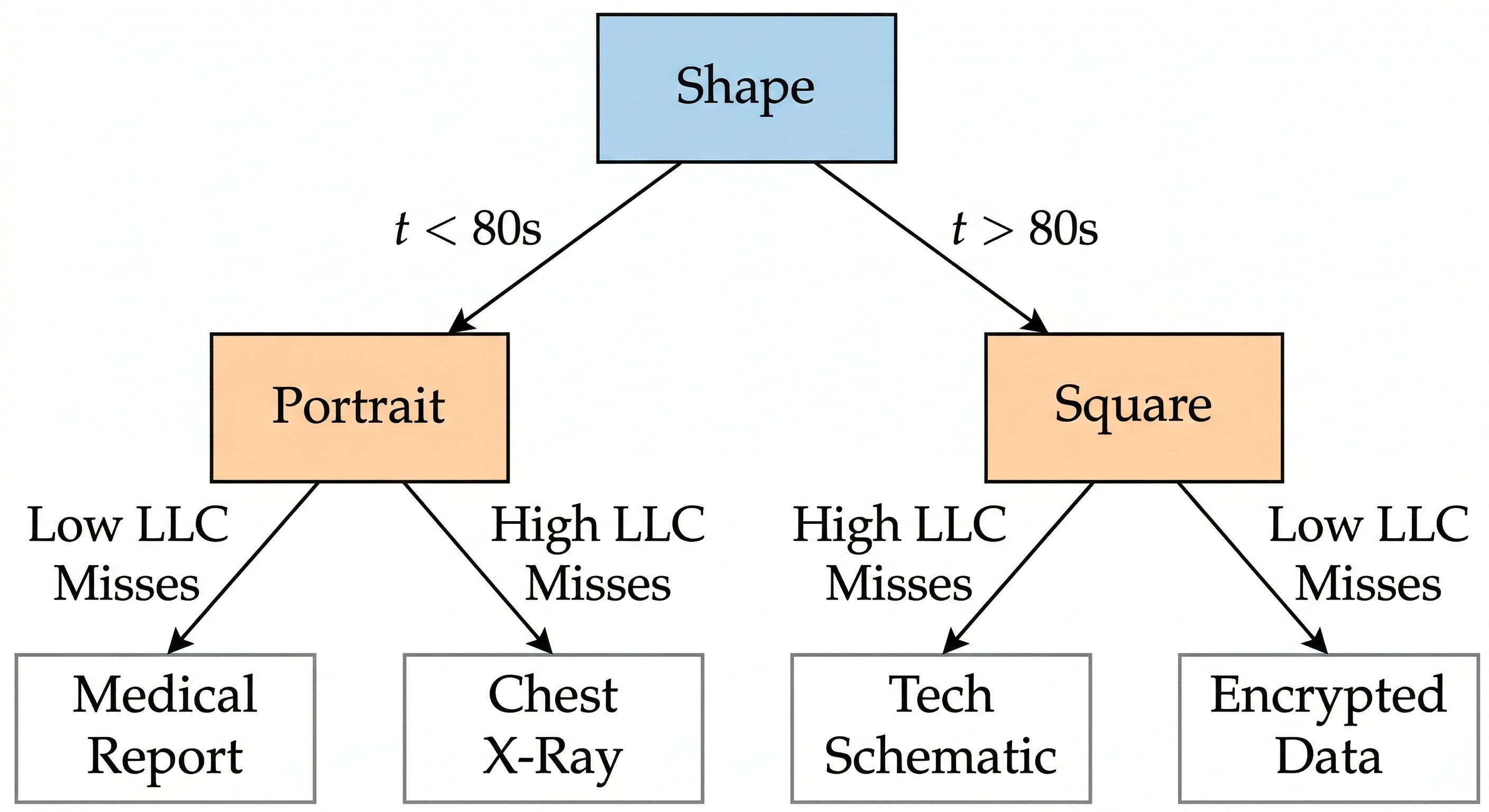}
    \caption{\textbf{The Physical Logic of the Attack.} The extracted decision tree perfectly mirrors the proposed dual-layer leakage model. \textbf{Layer 1 (Shape):} A fully deterministic root split based on inference time (e.g., $t < 80s$ vs. $t > 80s$) isolates the input's geometric grid into Portrait or Square. \textbf{Layer 2 (Substance):} Within each geometry cluster, cache contention levels (Low vs. High LLC Misses) differentiate inputs by visual density. This allows the attacker to reliably isolate dense structures (Chest X-Rays, Tech Schematics) from sparse ones (Medical Reports, Encrypted Data).}
    \label{fig:decision_tree}
\end{figure}

Interestingly, the model's decision tree (Figure \ref{fig:decision_tree}) closely follows the physical structure of our threat model: it first splits the data by inference time (capturing the image shape). Then it uses cache activity to refine the prediction based on visual content.

Most importantly, as detailed in Table~\ref{tab:classification_report}, the model achieves the highest recall for the most sensitive image categories, correctly identifying 100\% of Encrypted Data and 93\% of Chest X-rays. This means that even if some less structured documents are occasionally misclassified, an attacker can still reliably infer highly sensitive visual contexts from the execution of a local VLM.

\begin{table}[t]
    \centering
    \small
    \renewcommand{\arraystretch}{1.2}
    \begin{tabular}{l | c c c}
        \toprule
        \textbf{Class} & \textbf{Precision} & \textbf{Recall} & \textbf{F1-Score} \\
        \midrule
        Medical Report (Portrait) & 0.92 & 0.73 & 0.81 \\
        \textbf{Chest X-Ray (Portrait)} & 0.78 & \textbf{0.93} & 0.85 \\
        \textbf{Encrypted Data (Square)} & 0.79 & \textbf{1.00} & 0.88 \\
        Tech Schematic (Square) & 1.00 & 0.67 & 0.80 \\
        \midrule
        \textbf{Overall Accuracy} & \multicolumn{3}{c}{\textbf{84.0\%}} \\
        \bottomrule
    \end{tabular}
    \vspace{3pt}
    \caption{\textbf{Classification Report for the Combined Attack.} While achieving an overall accuracy of 84.0\%, the attack demonstrates asymmetric effectiveness. It achieves perfect (1.00) and near-perfect (0.93) recall for privacy-critical targets (Encrypted Data and Chest X-Rays). This indicates that the side-channel is highly reliable at identifying sensitive, visually dense inputs, ensuring they are rarely missed despite microarchitectural noise.}
    \label{tab:classification_report}
\end{table}

\section{Evaluation and Generalization}
\label{sec:evaluation}
Having introduced the attack vectors, we now evaluate the generality and robustness of the observed leakage across different VLM architectures, hardware platforms, and background system conditions.

\subsection{Cross-Model Validation}
To evaluate whether the observed side-channel is specific to a single implementation or reflects a broader design pattern, we compare LLaVA v1.6 against LLaVA v1.5 (static preprocessing) and Qwen2-VL (dynamic preprocessing). To execute this comparison, we developed an automated profiling framework that processes our geometric benchmark dataset through each model. The script directs the inference engine and extracts the prompt evaluation latency per image, systematically mapping input aspect ratios to their resulting execution times.

\begin{figure}[t]
    \centering
    \includegraphics[width=\linewidth]{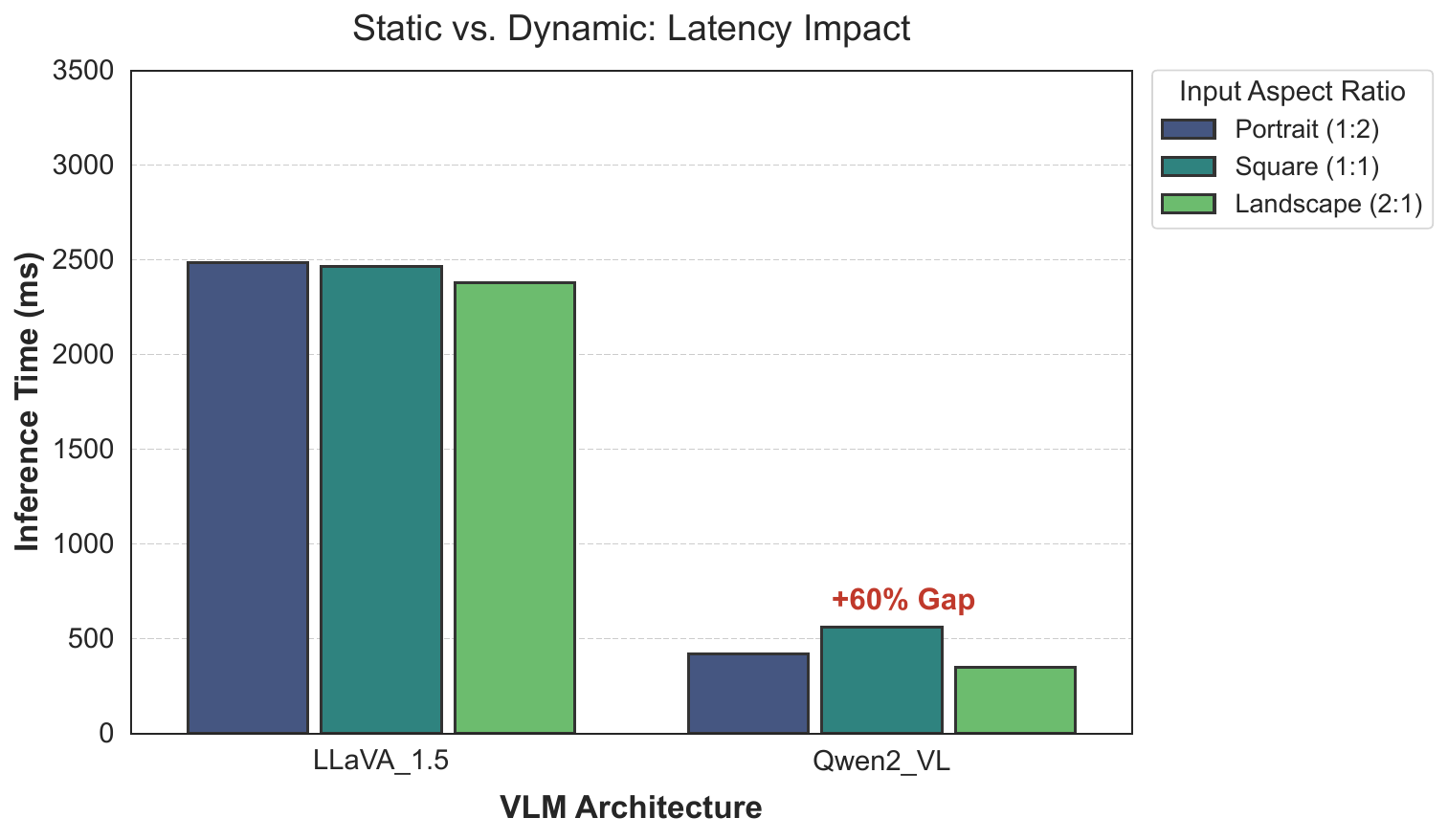}
    \caption{\textbf{Cross-Model Comparison.} LLaVA v1.5 (static preprocessing) shows similar latency across aspect ratios, whereas Qwen2-VL (dynamic preprocessing) exhibits a clear step-like timing pattern. Square inputs ($1:1$) incur approximately a 60\% latency increase compared to rectangular inputs.}
    \label{fig:cross_model}
\end{figure}

As shown in Figure~\ref{fig:cross_model}, the static model exhibits a nearly constant latency across aspect ratios. In contrast, Qwen2-VL reproduces the step-like timing behavior observed in our primary experiments, where square inputs incur an approximately $60\%$ latency increase compared to rectangular inputs. This consistency across independently developed models indicates that the leakage originates from the dynamic preprocessing algorithm itself. Because the number of processed patches depends on the input grid ($m \times n$), different aspect ratios lead to different control-flow paths, thereby directly affecting inference time. Consequently, the side-channel derives from the fundamental algorithmic design rather than specific model weights or implementation artifacts.

\subsection{Cross-Architecture Validation}
While the previous results established the attack's efficacy on Intel, we now evaluate the influence of cache capacity using an AMD-based system.

We replicated the attack on an AMD Ryzen 9 7950X processor, which has a relatively large LLC (Table~\ref{tab:specs}). The experiments were conducted using a Python-based framework that runs pre-compiled \texttt{llama.cpp} binaries on both the Intel and AMD systems. For each image in the benchmark, the script pins the inference process to a specific CPU core using \texttt{taskset} and collects hardware measurements using the Linux \texttt{perf} tool. In particular, we record LLC misses and execution cycles. This allows us to test whether the observed leakage depends on specific hardware properties or the algorithmic design.

We expect the geometric leakage, which appears as timing differences due to variable-grid processing, to remain visible regardless of cache size, as it is caused by input-dependent control flow in the algorithm. In contrast, cache-based signals such as LLC misses may vary with cache size, because a larger cache can reduce memory contention and thereby affect the number of misses. Table~\ref{tab:cross_arch_results} summarizes the results.

As expected, the cross-architecture experiment reveals two different types of leakage: one that remains stable across hardware platforms, and another that depends on microarchitectural characteristics.

\begin{enumerate}
    \item \textbf{Architecture-Independent Leakage (Aspect Ratio):} The timing-based side-channel remains visible on both platforms. Although the AMD processor runs faster overall, the relative latency gap between square and portrait inputs remains clear ($1.61\times$). This suggests that the leakage originates from the input-dependent control flow of the preprocessing algorithm rather than from hardware-specific behavior.
    
    \item \textbf{Hardware-Sensitive Leakage (Semantics):} In contrast, semantic leakage is more dependent on hardware characteristics. While clearly observable on the Intel platform, the signal is weaker on the AMD system, presumably due to its larger 64\,MB LLC. The larger cache reduces memory contention and dampens the observable cache-miss signal.
\end{enumerate}

\subsection{Robustness to System Load}
Finally, we evaluated the robustness of the semantic signal under realistic background system load. During each experiment, the VLM inference was executed concurrently with a synthetic workload generated using \texttt{stress-ng}. The stress process was configured to occupy two additional CPU cores and introduce sustained CPU and cache pressure (using the \texttt{--cpu} and \texttt{--l3-cache} stressors). By measuring LLC contention under these conditions, we assess whether the semantic signal remains observable despite continuous background cache activity (detailed results are provided in Appendix C of the supplementary material).

\section{Discussion and Design Recommendations}
\label{sec:discussion}

Our findings highlight a fundamental trade-off in local AI systems between architectural efficiency and side-channel resilience. To address practical security engineering concerns, we map the identified vulnerabilities to actionable design recommendations and evaluate the trade-offs of implementing them.

\subsection{Mitigation Strategies}
Effective mitigation requires addressing the two tiers of our threat model (Section \ref{sec:threat_model}) separately:
\begin{itemize}
    \item \textbf{Tier 1 (Timing/Geometry):} Mitigations must break the deterministic link between the input's aspect ratio and the total execution time. This requires application-level or framework-level interventions.
    \item \textbf{Tier 2 (Cache/Semantics):} Mitigations must hide the memory access patterns caused by the structural complexity of the input. This typically requires OS-level or hardware-level controls.
\end{itemize}

\subsection{Mitigating Tier 1: Application and Framework Interventions}
To eliminate geometric leakage, the VLM must perform a constant amount of computational work regardless of the input shape.

\subsubsection{Constant-Work Padding}
A straightforward mitigation is to enforce constant computational work by padding all inputs to the worst-case grid configuration (e.g., forcing a $2 \times 2$ grid that yields 5 patches). This approach removes the input-dependent control flow that enables the timing side-channel, as every input is processed using the same computational path. However, this mitigation incurs a substantial performance cost. On our reference Intel platform (Table~\ref{tab:cross_arch_results}), processing a portrait image naturally requires 49 seconds, while the worst-case square configuration requires 111 seconds. Enforcing constant-work padding would therefore introduce an approximate 126\% increase in inference latency for portrait inputs.

\subsubsection{Static Resolution ("Privacy Mode")}
As a lower-overhead alternative, local inference frameworks (e.g., \texttt{llama.cpp} or PyTorch) could provide an explicit ``Privacy Mode'' flag (e.g., \texttt{--disable-anyres}). In this mode, the model would bypass dynamic resolution entirely and revert to a static single-patch resizing strategy similar to earlier architectures such as LLaVA v1.5. While this reduces high-resolution visual fidelity, it avoids the substantial performance overhead of constant-work padding. It provides a simple baseline defense for privacy-sensitive applications, such as medical or document-processing workloads.

\subsection{Mitigating Tier 2: System-Level Interventions}
Unlike execution-time leakage, semantic cache leakage cannot be mitigated directly at the application level. As a result, effective defenses must rely on system-level controls.

\subsubsection{OS Telemetry Policies} 
The most immediate defense against Tier 2 is restricting access to hardware performance counters. Operating system vendors can adopt stricter defaults (e.g., setting \texttt{perf\_event\_paranoid = 3} on Linux) to prevent unprivileged applications from profiling shared caches during sensitive workloads. 

\subsubsection{Telemetry Obfuscation via Noise Injection} 
If telemetry access must remain enabled for debugging, OS schedulers could intentionally inject controlled background memory activity during VLM inference to lower the signal-to-noise ratio. For example, the scheduler could periodically trigger lightweight memory accesses or cache-thrashing routines on idle cores during inference. However, this trade-off reduces overall system throughput and negatively impacts battery life on mobile devices.

\subsection{The Hardware Lottery}
Finally, our cross-architecture evaluation suggests a form of "hardware lottery" \cite{hooker2020hardware}. Users on consumer-grade hardware with modest LLC capacities (e.g., 30 MB) exhibit observable semantic leakage. In contrast, high-end systems with large shared caches (e.g., 64 MB) incidentally absorb the working set variance, naturally dampening the Tier 2 signal. This asymmetry highlights a critical risk for future deployments. As VLMs transition to specialized, resource-constrained edge hardware, these devices will inherently have significantly smaller LLC capacities. Consequently, the dampening effect observed on large desktop caches will vanish, making semantic side-channel leakages even more prominent. Evaluating microarchitectural vulnerability must therefore become a mandatory security benchmark for dedicated Edge AI hardware.

\section{Security Implications and Limitations}
\label{sec:implications_limitations}

While our evaluation identifies a fundamental architectural vulnerability, assessing its real-world impact requires considering how these systems are deployed in practice and recognizing the limitations of our current analysis.

\subsection{Real-World Risk Scenarios}
The shift towards local VLMs is largely driven by industries handling highly sensitive data. The dual-layer side-channel exposes these deployments to significant privacy risks:
\begin{itemize}
    \item \textbf{Medical Diagnostics:} A local VLM deployed in a clinic might process both standard patient records (sparse, portrait documents) and diagnostic imaging (dense, square or landscape X-rays). An unprivileged attacker can easily distinguish between these tasks using our combined attack, violating patient confidentiality without explicitly accessing the protected data.
    \item \textbf{Enterprise and Legal Operations:} Organizations often deploy local AI to review confidential contracts or intellectual property. The deterministic geometric leakage (Tier 1) allows an attacker to reliably distinguish when an employee is reviewing standard legal documents versus specialized technical documents, potentially revealing sensitive patterns of corporate activity.
\end{itemize}

\subsection{Limitations and Future Work}
Our experimental methodology was designed to optimally isolate and prove the existence of algorithmic side-channels in dynamic preprocessing. Building on this foundation, we identify several key directions for future research:

\subsubsection{Isolating the Algorithmic Signal}
Our experiments demonstrate the presence of the vulnerability under a consistent experimental setup. Future work could further examine how system-level conditions influence the strength of the leakage. In particular, studies could repeat the same visual workloads while varying common hardware settings, such as CPU power-management policies, the use of Simultaneous Multithreading (SMT), or the memory layout in multi-socket systems (NUMA). Such experiments would help clarify how much of the observed signal is driven by the dynamic grid mechanism itself versus broader system configuration effects.

\subsubsection{Impact of Cache Capacity} 
Our cross-architecture evaluation demonstrates that LLC capacity influences the semantic (Tier 2) leakage. Future work should systematically quantify this relationship across a broader spectrum of processors to establish the exact cache thresholds at which semantic variance becomes unobservable.

\subsubsection{Algorithmic Vulnerability in Edge AI Devices}
Like all cache-based attacks, our semantic signal can be disrupted by extreme memory contention. However, our primary contribution is exposing the underlying algorithmic vulnerability: the VLM's attention mechanism inherently alters its memory footprint based on visual entropy. Since this algorithmic behavior manifests physically depending on the specific hardware architecture, evaluating resilience against this leakage vector should become a mandatory, device-specific security benchmark for all future dedicated Edge AI hardware.

\section{Conclusion}
\label{sec:conclusion}
The primary motivation for deploying Edge AI and local VLMs is to preserve data privacy. However, we have shown that the architectural shift toward dynamic high-resolution preprocessing introduces a previously overlooked multi-dimensional side-channel. By jointly exploiting deterministic geometric leakage (Tier 1) through execution time and probabilistic semantic leakage (Tier 2) through cache behavior, an unprivileged attacker can infer not only the shape but also high-level contextual properties of visual inputs.

Our evaluation across architectures and system conditions indicates that the geometric component of this leakage stems from the algorithmic design of dynamic preprocessing. In contrast, semantic leakage is influenced by underlying microarchitectural characteristics such as cache capacity.

More broadly, our results highlight that local execution alone does not guarantee privacy on shared hardware. Future edge AI systems must therefore treat side-channel resilience as a first-class design consideration rather than assuming that on-device inference inherently provides privacy.

\section*{Acknowledgment}
To facilitate reproducibility, the experimental measurement scripts, feature extraction tools, and a representative sample of the dataset have been made publicly available at: \url{https://anonymous.4open.science/r/Shape-and-Substance-Dual-Layer-Side-Channel-Attacks-on-Local-Vision-Language-Models-C6D7}. 

\bibliographystyle{IEEEtran}
\bibliography{references}

@inproceedings{radford2021learning,
  title={Learning transferable visual models from natural language supervision},
  author={Radford, Alec and Kim, Jong Wook and Hallacy, Chris and Ramesh, Aditya and Goh, Gabriel and Agarwal, Sandhini and Sastry, Girish and Askell, Amanda and Mishkin, Pamela and Clark, Jack and others},
  booktitle={International conference on machine learning},
  pages={8748--8763},
  year={2021},
  organization={PMLR}
}

@inproceedings{liu2024improved,
  title={Improved baselines with visual instruction tuning},
  author={Liu, Haotian and Li, Chunyuan and Li, Yuheng and Lee, Yong Jae},
  booktitle={Proceedings of the IEEE/CVF conference on computer vision and pattern recognition},
  pages={26296--26306},
  year={2024}
}

@article{li2024llava,
  title={Llava-next-interleave: Tackling multi-image, video, and 3d in large multimodal models},
  author={Li, Feng and Zhang, Renrui and Zhang, Hao and Zhang, Yuanhan and Li, Bo and Li, Wei and Ma, Zejun and Li, Chunyuan},
  journal={arXiv preprint arXiv:2407.07895},
  year={2024}
}

@article{wang2024qwen2,
  title={Qwen2-vl: Enhancing vision-language model's perception of the world at any resolution},
  author={Wang, Peng and Bai, Shuai and Tan, Sinan and Wang, Shijie and Fan, Zhihao and Bai, Jinze and Chen, Keqin and Liu, Xuejing and Wang, Jialin and Ge, Wenbin and others},
  journal={arXiv preprint arXiv:2409.12191},
  year={2024}
}

@article{li2019edge,
  title={Edge AI: On-demand accelerating deep neural network inference via edge computing},
  author={Li, En and Zeng, Liekang and Zhou, Zhi and Chen, Xu},
  journal={IEEE transactions on wireless communications},
  volume={19},
  number={1},
  pages={447--457},
  year={2019},
  publisher={IEEE}
}

@inproceedings{carlini2021extracting,
  title={Extracting Training Data from Large Language Models},
  author={Carlini, Nicholas and Tramer, Florian and Wallace, Eric and Jagielski, Matthew and Herbert-Voss, Ariel and Lee, Katherine and Roberts, Adam and Brown, Tom and Song, Dawn and Erlingsson, Ulfar and others},
  booktitle={30th USENIX Security Symposium (USENIX Security 21)},
  pages={2633--2650},
  year={2021}
}

@inproceedings{ristenpart2009hey,
  title={Hey, you, get off of my cloud: exploring information leakage in third-party compute clouds},
  author={Ristenpart, Thomas and Tromer, Eran and Shacham, Hovav and Savage, Stefan},
  booktitle={Proceedings of the 16th ACM Conference on Computer and Communications Security (CCS)},
  pages={199--212},
  year={2009}
}

@inproceedings{naghibijouybari2018rendered,
  title={Rendered Insecure: GPU Side Channel Attacks are Practical},
  author={Naghibijouybari, Hoda and Neupane, Ajaya and Qian, Zhiyun and Abu-Ghazaleh, Nael},
  booktitle={27th USENIX Security Symposium (USENIX Security 18)},
  pages={2139--2156},
  year={2018}
}

@inproceedings{yarom2014flush,
  title={FLUSH+ RELOAD: A High Resolution, Low Noise, L3 Cache Side-Channel Attack},
  author={Yarom, Yuval and Falkner, Katrina},
  booktitle={23rd USENIX Security Symposium (USENIX Security 14)},
  pages={719--732},
  year={2014}
}

@misc{llamacpp,
  author = {Georgi Gerganov and contributors},
  title  = {llama.cpp},
  year   = {2024},
  howpublished = {\url{https://github.com/ggml-org/llama.cpp}},
}

@article{ge2018survey,
  title={A survey of microarchitectural timing attacks and countermeasures on contemporary hardware},
  author={Ge, Qian and Yarom, Yuval and Cock, David and Heiser, Gernot},
  journal={Journal of Cryptographic Engineering},
  volume={8},
  number={1},
  pages={1--27},
  year={2018},
  publisher={Springer}
}

@article{tschannen2025siglip,
  title={Siglip 2: Multilingual vision-language encoders with improved semantic understanding, localization, and dense features},
  author={Tschannen, Michael and Gritsenko, Alexey and Wang, Xiao and Naeem, Muhammad Ferjad and Alabdulmohsin, Ibrahim and Parthasarathy, Nikhil and Evans, Talfan and Beyer, Lucas and Xia, Ye and Mustafa, Basil and others},
  journal={arXiv preprint arXiv:2502.14786},
  year={2025}
}

@inproceedings{li2023blip,
  title={Blip-2: Bootstrapping language-image pre-training with frozen image encoders and large language models},
  author={Li, Junnan and Li, Dongxu and Savarese, Silvio and Hoi, Steven},
  booktitle={International conference on machine learning},
  pages={19730--19742},
  year={2023},
  organization={PMLR}
}

@article{alayrac2022flamingo,
  title={Flamingo: a visual language model for few-shot learning},
  author={Alayrac, Jean-Baptiste and Donahue, Jeff and Luc, Pauline and Miech, Antoine and Barr, Iain and Hasson, Yana and Lenc, Karel and Mensch, Arthur and Millican, Katherine and Reynolds, Malcolm and others},
  journal={Advances in neural information processing systems},
  volume={35},
  pages={23716--23736},
  year={2022}
}

@article{gao2025know,
  title={I Know What You Said: Unveiling Hardware Cache Side-Channels in Local Large Language Model Inference},
  author={Gao, Zibo and Hu, Junjie and Guo, Feng and Zhang, Yixin and Han, Yinglong and Liu, Siyuan and Li, Haiyang and Lv, Zhiqiang},
  journal={arXiv preprint arXiv:2505.06738},
  year={2025}
}

@article{zheng2024inputsnatch,
  title={Inputsnatch: Stealing input in llm services via timing side-channel attacks},
  author={Zheng, Xinyao and Han, Husheng and Shi, Shangyi and Fang, Qiyan and Du, Zidong and Hu, Xing and Guo, Qi},
  journal={arXiv preprint arXiv:2411.18191},
  year={2024}
}

@article{song2025early,
  title={The early bird catches the leak: Unveiling timing side channels in llm serving systems},
  author={Song, Linke and Pang, Zixuan and Wang, Wenhao and Wang, Zihao and Wang, XiaoFeng and Chen, Hongbo and Song, Wei and Jin, Yier and Meng, Dan and Hou, Rui},
  journal={IEEE Transactions on Information Forensics and Security},
  year={2025},
  publisher={IEEE}
}

@article{gu2025auditing,
  title={Auditing prompt caching in language model apis},
  author={Gu, Chenchen and Li, Xiang Lisa and Kuditipudi, Rohith and Liang, Percy and Hashimoto, Tatsunori},
  journal={arXiv preprint arXiv:2502.07776},
  year={2025}
}

@inproceedings{ding2025moecho,
  title={MoEcho: Exploiting Side-Channel Attacks to Compromise User Privacy in Mixture-of-Experts LLMs},
  author={Ding, Ruyi and Xu, Tianhong and Shen, Xinyi and Ding, Aidong Adam and Fei, Yunsi},
  booktitle={Proceedings of the 2025 ACM SIGSAC Conference on Computer and Communications Security},
  pages={2159--2173},
  year={2025}
}

@article{adiletta2025spill,
  title={Spill The Beans: Exploiting CPU Cache Side-Channels to Leak Tokens from Large Language Models},
  author={Adiletta, Andrew and Sunar, Berk},
  journal={arXiv preprint arXiv:2505.00817},
  year={2025}
}

@inproceedings{wang2019keystrokes,
  title={Unveiling Your Keystrokes: A Cache-Based Side-Channel Attack on Graphics Libraries},
  author={Wang, Daimeng and Neupane, Ajaya and Qian, Zhiyun and Abu-Ghazaleh, Nael},
  booktitle={NDSS},
  year={2019}
}

@inproceedings{pessl2016drama,
  title={$\{$DRAMA$\}$: Exploiting $\{$DRAM$\}$ addressing for $\{$Cross-CPU$\}$ attacks},
  author={Pessl, Peter and Gruss, Daniel and Maurice, Cl{\'e}mentine and Schwarz, Michael and Mangard, Stefan},
  booktitle={25th USENIX security symposium (USENIX security 16)},
  pages={565--581},
  year={2016}
}

@inproceedings{shokri2017membership,
  title={Membership inference attacks against machine learning models},
  author={Shokri, Reza and Stronati, Marco and Song, Congzheng and Shmatikov, Vitaly},
  booktitle={2017 IEEE symposium on security and privacy (SP)},
  pages={3--18},
  year={2017},
  organization={IEEE}
}

@article{zhou2019edge,
  title={Edge intelligence: Paving the last-mile of artificial intelligence with edge computing},
  author={Zhou, Zhi and Chen, Xu and Li, En and Zeng, Liekang and Luo, Ke and Zhang, Junshan},
  journal={Proceedings of the IEEE},
  volume={107},
  number={8},
  pages={1738--1762},
  year={2019},
  publisher={IEEE}
}

@inproceedings{dosovitskiy2021image,
  title={An Image is Worth 16x16 Words: Transformers for Image Recognition at Scale},
  author={Dosovitskiy, Alexey and Beyer, Lucas and Kolesnikov, Alexander and Weissenborn, Dirk and Zhai, Xiaohua and Unterthiner, Thomas and Dehghani, Mostafa and Minderer, Matthias and Heigold, Georg and Gelly, Sylvain and others},
  booktitle={International Conference on Learning Representations},
  year={2021}
}

@inproceedings{vaswani2017attention,
  title={Attention is all you need},
  author={Vaswani, Ashish and Shazeer, Noam and Parmar, Niki and Uszkoreit, Jakob and Jones, Llion and Gomez, Aidan N and Kaiser, {\L}ukasz and Polosukhin, Illia},
  booktitle={Advances in neural information processing systems},
  pages={5998--6008},
  year={2017}
}

@inproceedings{shacham2004effectiveness,
  title={On the effectiveness of address-space randomization},
  author={Shacham, Hovav and Page, Matthew and Pfaff, Ben and Goh, Eu-Jin and Modadugu, Nagendra and Boneh, Dan},
  booktitle={Proceedings of the 11th ACM conference on Computer and communications security},
  pages={298--307},
  year={2004}
}

@inproceedings{liu2015last,
  title={Last-level cache side-channel attacks are practical},
  author={Liu, Fangfei and Yarom, Yuval and Ge, Qian and Heiser, Gernot and Lee, Ruby B},
  booktitle={2015 IEEE symposium on security and privacy},
  pages={605--622},
  year={2015},
  organization={IEEE}
}

@inproceedings{kocher2019spectre,
  title={Spectre attacks: Exploiting speculative execution},
  author={Kocher, Paul and Horn, Jann and Fogh, Anders and Genkin, Daniel and Gruss, Daniel and Haas, Werner and Hamburg, Mike and Lipp, Moritz and Mangard, Stefan and Prescher, Thomas and others},
  booktitle={2019 IEEE Symposium on Security and Privacy (SP)},
  pages={1--19},
  year={2019},
  organization={IEEE}
}

@article{schwartz2020green,
  title={Green AI},
  author={Schwartz, Roy and Dodge, Jesse and Smith, Noah A and Etzioni, Oren},
  journal={Communications of the ACM},
  volume={63},
  number={12},
  pages={54--63},
  year={2020},
  publisher={ACM New York, NY, USA}
}

@inproceedings{he2016deep,
  title={Deep residual learning for image recognition},
  author={He, Kaiming and Zhang, Xiangyu and Ren, Shaoqing and Sun, Jian},
  booktitle={Proceedings of the IEEE conference on computer vision and pattern recognition},
  pages={770--778},
  year={2016}
}

@inproceedings{schuster2017beauty,
  title={Beauty and the Burst: Remote Identification of Encrypted Video Streams},
  author={Schuster, Roei and Shmatikov, Vitaly and Tromer, Eran},
  booktitle={26th USENIX Security Symposium (USENIX Security 17)},
  pages={1357--1374},
  year={2017}
}

@inproceedings{cai2012touching,
  title={Touching from a distance: Website fingerprinting attacks and defenses},
  author={Cai, Xiang and Zhang, Xin Cheng and Joshi, Brijesh and Johnson, Rob},
  booktitle={Proceedings of the 2012 ACM conference on Computer and communications security},
  pages={605--616},
  year={2012}
}

@article{achiam2023gpt,
  title={GPT-4 Technical Report},
  author={Achiam, Josh and others},
  journal={arXiv preprint arXiv:2303.08774},
  year={2023}
}

@inproceedings{dao2022flash,
  title={FlashAttention: Fast and Memory-Efficient Exact Attention with IO-Awareness},
  author={Dao, Tri and Fu, Daniel Y and Ermon, Stefano and Rudra, Atri and R{\'e}, Christopher},
  booktitle={Advances in Neural Information Processing Systems},
  volume={35},
  pages={16344--16359},
  year={2022}
}

@inproceedings{dehghani2024patch,
  title={Patch n' Pack: NaViT, a Vision Transformer for any Aspect Ratio and Resolution},
  author={Dehghani, Mostafa and Mustafa, Basil and Djolonga, Josip and Heek, Jonathan and Gilmer, Justin and others},
  booktitle={Advances in Neural Information Processing Systems (NeurIPS)},
  volume={36},
  year={2024}
}

@inproceedings{crosby2003denial,
  title={Denial of Service via Algorithmic Complexity Attacks},
  author={Crosby, Scott A and Wallach, Dan S},
  booktitle={12th USENIX Security Symposium (USENIX Security 03)},
  pages={29--44},
  year={2003}
}

@inproceedings{yan2020cache,
  title={Cache Telepathy: Leveraging Shared Resource Attacks to Learn DNN Architectures},
  author={Yan, Mengjia and Fletcher, Christopher W and Torrellas, Josep},
  booktitle={29th USENIX Security Symposium (USENIX Security 20)},
  pages={2003--2020},
  year={2020}
}

@inproceedings{kotcher2013cross,
  title={Cross-origin pixel stealing: timing attacks using CSS filters},
  author={Kotcher, Robert and Pei, Yutong and Jumde, Pranjal and Jackson, Collin},
  booktitle={20th ACM Conference on Computer and Communications Security (CCS)},
  pages={1055--1062},
  year={2013}
}

@inproceedings{fredrikson2015model,
  title={Model inversion attacks that exploit confidence information and basic countermeasures},
  author={Fredrikson, Matt and Jha, Somesh and Ristenpart, Thomas},
  booktitle={Proceedings of the 22nd ACM SIGSAC Conference on Computer and Communications Security (CCS)},
  pages={1322--1333},
  year={2015}
}

@inproceedings{ribeiro2016should,
  title={"Why should I trust you?" Explaining the predictions of any classifier},
  author={Ribeiro, Marco Tulio and Singh, Sameer and Guestrin, Carlos},
  booktitle={Proceedings of the 22nd ACM SIGKDD international conference on knowledge discovery and data mining},
  pages={1135--1144},
  year={2016}
}

@article{hooker2020hardware,
  title={The hardware lottery},
  author={Hooker, Sara},
  journal={Communications of the ACM},
  volume={64},
  number={12},
  pages={58--65},
  year={2021},
  publisher={ACM New York, NY, USA}
}

@inproceedings{kocher1996timing,
  title={Timing attacks on implementations of Diffie-Hellman, RSA, DSS, and other systems},
  author={Kocher, Paul C},
  booktitle={Annual international cryptology conference},
  pages={104--113},
  year={1996},
  organization={Springer}
}

@article{zhang2023guessing,
title={A guessing entropy-based framework for deep learning-assisted side-channel analysis},
author={Zhang, Ziyue and Ding, A Adam and Fei, Yunsi},
journal={IEEE Transactions on Information Forensics and Security},
volume={18},
pages={3018--3030},
year={2023},
publisher={IEEE}
}

@article{zhang2021stealing,
title={Stealing neural network structure through remote FPGA side-channel analysis},
author={Zhang, Yicheng and Yasaei, Rozhin and Chen, Hao and Li, Zhou and Al Faruque, Mohammad Abdullah},
journal={IEEE Transactions on Information Forensics and Security},
volume={16},
pages={4377--4388},
year={2021},
publisher={IEEE}
}

@article{kim2023deep,
title={Deep learning-based detection for multiple cache side-channel attacks},
author={Kim, Hodong and Hahn, Changhee and Kim, Hyunwoo J and Shin, Youngjoo and Hur, Junbeom},
journal={IEEE Transactions on Information Forensics and Security},
volume={19},
pages={1672--1686},
year={2023},
publisher={IEEE}
}


\setcounter{figure}{0}
\setcounter{table}{0}
\setcounter{equation}{0}
\renewcommand{\thefigure}{S\arabic{figure}}
\renewcommand{\thetable}{S\arabic{table}}
\renewcommand{\theequation}{S\arabic{equation}}
\counterwithin*{figure}{section}
\counterwithin*{table}{section}
\counterwithin*{equation}{section}

\appendices

\section{Artifact Reproducibility Details}
\label{app:reproducibility}

To facilitate exact reproduction of the reported side-channel leakage, we provide the specific configuration parameters used in our evaluation.

\paragraph{Victim \& Attacker Isolation}
\textbf{Process Pinning for Experimental Control.}
To reduce scheduling-induced variance and enable reproducible attribution of observed leakage to shared microarchitectural resources, we pinned the victim and attacker processes to specific CPU cores using \texttt{taskset}.
We use core pinning to isolate the microarchitectural signal from OS scheduler noise, thereby establishing a precise baseline for algorithmic leakage. While an unprivileged attacker cannot dictate the victim's scheduling, they can monitor core utilization and dynamically control their own CPU affinity (e.g., via \texttt{sched\_setaffinity}) to co-locate with the target process actively. Thus, our pinned setup evaluates the vulnerability under optimal conditions of shared-cache contention, which a realistic attacker actively strives to achieve.

Specifically, we used the following configuration:
\begin{itemize}
    \item \textbf{Victim (VLM):} Pinned to physical core 4 (\texttt{taskset -c 4}).
    \item \textbf{Attacker (Profiler):} Pinned either to physical core 5 (to induce contention at the shared LLC) or to the same physical core (to evaluate a hyperthreading setting), depending on the experiment described in Section IV of the main manuscript.
\end{itemize}

This setup allows us to isolate leakage arising from shared cache resources, either through L1/L2 sharing via simultaneous multi-threading or through contention in the shared L3 cache, while minimizing interference from OS scheduling.

\paragraph{Software Versions}
Our reference implementation is based on \texttt{llama.cpp}.
To ensure a consistent memory layout and stable execution paths across runs, all experiments were conducted using a fixed commit (\texttt{b3456...}).
We employed the \texttt{Q4\_K\_M} quantization scheme, which corresponds to the default configuration for high-performance local inference.

\paragraph{Kernel Configuration}
\textbf{Performance Counter Access for Ground Truth Collection.}
For the purpose of constructing a clean and reproducible ground-truth dataset, we temporarily configured
\texttt{kernel.perf\_event\_paranoid = -1} during data collection.
This setting was used exclusively to enable unrestricted access to performance counters for measurement validation.

Importantly, the attack itself does not rely on elevated privileges.
As described in Section IV of the main manuscript, the specific hardware events exploited in this work (e.g., LLC misses and CPU cycles) are accessible to unprivileged user-space processes under the default restrictive configuration
(\texttt{perf\_event\_paranoid} $\leq 2$), which is commonly used in standard Linux distributions such as Ubuntu and Debian.

\section{Visual Density \& Spectral Analysis}
\label{app:visuals}

In Section VI of the main manuscript, we argued that the observed probabilistic leakage is related to differences in spatial frequency and visual structure across image patches. This appendix provides visual examples that illustrate these low-level properties without relying on the image's semantic meaning.

\begin{figure}[h!]
    \centering

    \begin{minipage}{0.48\columnwidth}
        \centering
        \includegraphics[width=\linewidth]{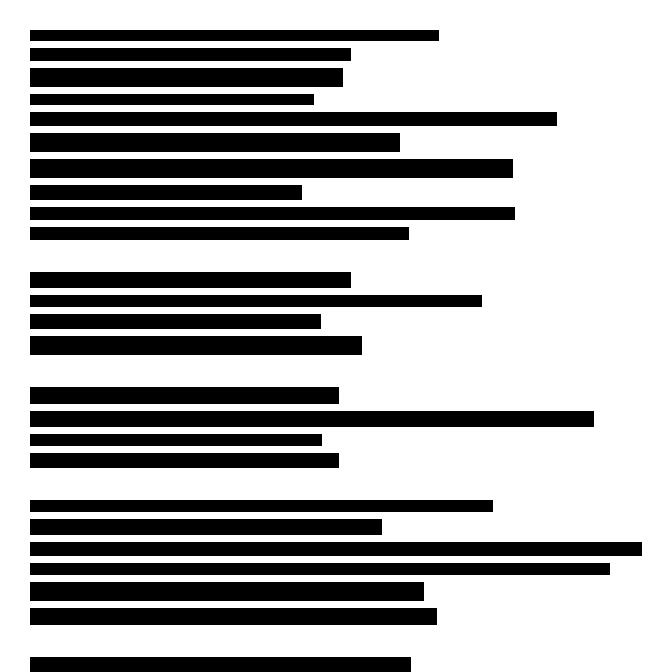} 
        \vspace{0.1cm}
        \scriptsize{(a) Document}
    \end{minipage}
    \hfill
    \begin{minipage}{0.48\columnwidth}
        \centering
        \includegraphics[width=\linewidth]{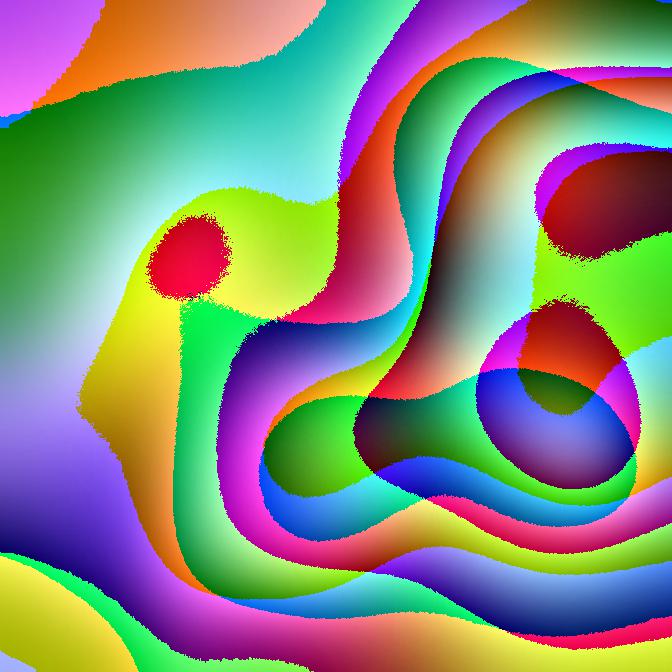} 
        \vspace{0.1cm}
        \scriptsize{(b) Nature}
    \end{minipage}
    
    \vspace{0.3cm}

    \begin{minipage}{0.48\columnwidth}
        \centering
        \includegraphics[width=\linewidth]{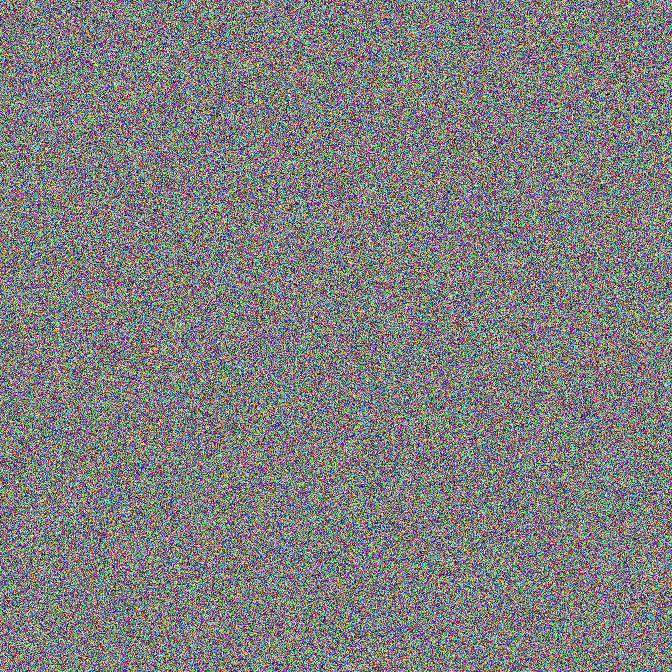} 
        \vspace{0.1cm}
        \scriptsize{(c) Noise}
    \end{minipage}
    \hfill
    \begin{minipage}{0.48\columnwidth}
        \centering
        \includegraphics[width=\linewidth]{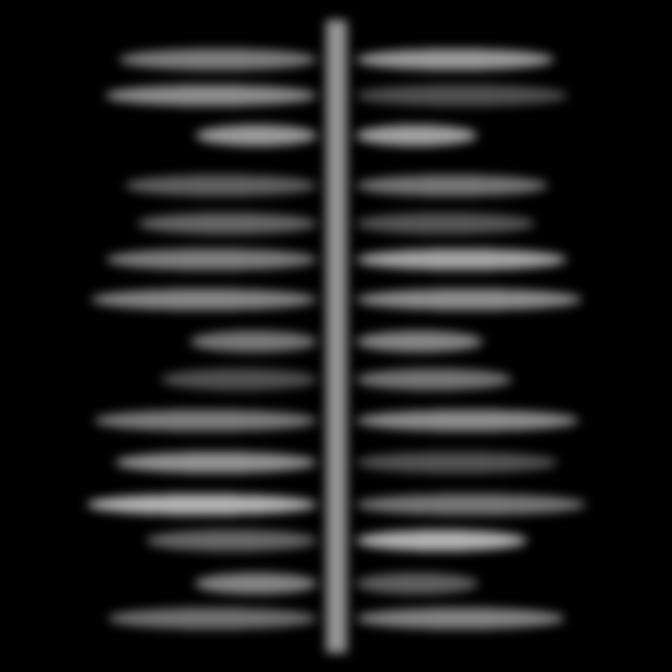} 
        \vspace{0.1cm}
        \scriptsize{(d) X-Ray}
    \end{minipage}
    
    \caption{\textbf{Visual Density Comparison.}
    Representative samples illustrating differences in spatial frequency and structural density.
    High-contention classes (Nature, X-Ray) exhibit dense textures across the image plane,
    while low-contention classes (Document, Noise) contain large homogeneous regions with limited fine-grained structure.}
    
    \label{fig:visual_density}
\end{figure}

Figure~\ref{fig:visual_density} illustrates representative samples from the four evaluated categories.
\begin{itemize}

\item \textbf{Low Contention (Left Column):}
Categories such as \textit{Documents} and \textit{Crypto-Noise} contain large uniform regions and relatively little fine texture. As a result, the vision encoder processes these inputs with more regular memory access patterns, thereby reducing LLC contention.

\item \textbf{High Contention (Right Column):}
Categories such as \textit{Nature} and \textit{X-Rays} contain dense visual structure and rich textures across the image. Processing these inputs activates additional parts of the vision encoder and increases the working set size, thereby leading to higher LLC contention.

\end{itemize}

\section{Robustness Under System Load}
\label{app:robustness}
\begin{table}[h!]
    \centering
    \small
    \renewcommand{\arraystretch}{1.2}
    \resizebox{\columnwidth}{!}{
    \begin{tabular}{l c c c}
        \toprule
        \textbf{Condition} & \textbf{Noise (Avg)} & \textbf{X-Ray (Avg)} & \textbf{Separation Gap} \\
        \midrule
        \textbf{Idle State}   & 16.9 & 17.9 & \textbf{1.0} \\
        \textbf{System Load} & 17.1 & 18.1 & \textbf{1.0} \\
        \bottomrule
    \end{tabular}
    }
    \caption{\textbf{Leakage Robustness.}
    Mean LLC misses ($\times 10^9$) under idle and background-load conditions.
    While absolute miss counts increase under load, the relative separation remains stable.}
    \label{tab:load_robustness}
\end{table}

A common critique of microarchitectural side-channels is their sensitivity to background system activity.
To evaluate robustness under realistic noise, we repeated the content-based inference experiment while subjecting the system to controlled synthetic load using \texttt{stress-ng}, saturating two additional CPU cores.

Table~\ref{tab:load_robustness} summarizes the results.
As expected, background load increases the absolute LLC miss counts for all inputs, raising the overall noise floor.
However, the \textbf{relative separation gap ($\Delta$)} between structurally dense inputs (X-Ray) and noise-like inputs (Crypto-Noise) remains stable at approximately $1.0 \times 10^9$ misses.

This observation indicates that moderate background load does not eliminate the cache-based separation between content classes.
The stability of the separation gap suggests that the vision encoder’s memory footprint during the prompt-evaluation phase dominates cache behavior over short execution intervals, rendering the leakage resilient to typical background activity.

\end{document}